\documentclass[aps,prl,twocolumn,superscriptaddress,showpacs]{revtex4-1}

\usepackage{amsmath,amssymb,graphics,epsfig,epstopdf,color,verbatim,tabularx}
\usepackage{graphicx}
\usepackage{color}

\DeclareGraphicsExtensions{.pdf,.png,.jpg}

\begin{document}

\title{Giant Resonant Self-Blocking in Magnetophononically Driven Quantum 
Magnets}

\author{M. Yarmohammadi}
\affiliation{Condensed Matter Theory, TU Dortmund University, Otto-Hahn-Strasse 
4, 44221 Dortmund, Germany}

\author{M. Krebs}
\affiliation{Condensed Matter Theory, TU Dortmund University, Otto-Hahn-Strasse 
4, 44221 Dortmund, Germany}

\author{G. S. Uhrig}
\affiliation{Condensed Matter Theory, TU Dortmund University, Otto-Hahn-Strasse 
4, 44221 Dortmund, Germany}

\author{B. Normand}
\affiliation{Paul Scherrer Institute, CH-5232 Villigen-PSI, Switzerland}
\affiliation{Institute of Physics, Ecole Polytechnique F\'{e}d\'{e}rale 
de Lausanne (EPFL), CH-1015 Lausanne, Switzerland}

\begin{abstract}
Magnetophononics, the modulation of magnetic interactions by driven 
infrared-active lattice excitations, is emerging as a key mechanism for the 
ultrafast dynamical control of both semiclassical and quantum spin systems 
by coherent light. We demonstrate that, in a quantum magnetic material with 
strong spin-phonon coupling, resonances between the driven phonon and the 
spin-excitation frequencies exhibit a giant self-blocking effect. Instead of 
absorbing more energy, the spin system acts as a strong brake on the driven 
phonon, causing it to absorb only a tiny fraction of the power available 
from the laser. Using the quantum master equations governing the 
nonequilibrium steady states of the coupled spin-lattice system, we show 
how self-blocking dominates the dynamics, demonstrate the creation of 
mutually repelling hybrid spin-phonon states, and control the nonequilibrium 
renormalization of the lattice-driven spin excitation band.
\end{abstract}


\maketitle


Rapid advances in laser technology \cite{Salen19} have made it possible not 
only to probe but also to pump quantum materials in a controlled manner on 
ultrafast timescales and at all the frequencies relevant to excitations in 
condensed matter \cite{rktn,rzsz,Buzzi18}. This has led to phenomena ranging 
from Floquet engineering of electronic band structures \cite{rok} to enhanced 
superconductivity \cite{rcr} and switching of the metal-insulator transition 
\cite{Iwai03}. A wide range of experimental and theoretical efforts is 
now under way to extend such ultrafast control to every aspect of strongly 
correlated materials beyond the charge, including lattice, orbital, spin, 
nematic, and chiral degrees of freedom \cite{rtkcgms}.

Among these, spin systems offer perhaps the ultimate quantum many-body states 
due to their intrinsically high entanglement and relatively low energy scales, 
which lead to rather clean experimental realizations. Ultrafast switching, 
modulation, transport, and destruction of semiclassical ordered magnetism 
have been achieved using light of different frequencies \cite{Kampfrath11,
Mikhaylovskiy15,rjbasyzb}. However, coupling to a magnetic order parameter is 
often not appropriate for the dynamical control of quantum magnetic materials, 
and increasing attention is focused on using the lattice as an intermediary 
\cite{Disa20,Afanasiev21,Deltenre21,Bossini21,Disa21}. While ``nonlinear 
phononics'' \cite{Foerst11} exploits the anharmonic lattice potential, to 
date for low-frequency magnetic control \cite{Nova17}, ``magnetophononics'' 
\cite{Fechner18} uses harmonic phonons to gain fully frequency-selective 
control of exchange-type interactions \cite{Giorgianni21}. Strong excitation 
of collective spin modes at their intrinsic frequencies opens the possibility 
not only of effecting the slow or fast (Floquet) modulation of existing 
magnetic states but also of creating fundamentally different types of hybrid 
state.

In this Letter we show that the magnetophononic mechanism has an intrinsic 
giant self-blocking effect, by which a driven phonon in resonance with the 
peak density of magnetic excitations absorbs very little of the driving 
laser power. We demonstrate self-blocking by considering the nonequilibrium 
steady states (NESS) of an alternating quantum spin chain strongly coupled 
to a bulk Einstein phonon mode. We compute the driving-induced mutual 
renormalization of the lattice and spin excitations, finding that 
distinctive hybrid excitations emerge for phonon frequencies near the 
spin-band edges and that other in-band frequencies effect a global reshaping 
of the spin spectrum. We discuss the consequences of self-blocking and 
dynamical spectral renormalization for pump-probe experiments on quantum 
magnetic materials such as CuGeO$_3$ and (VO)$_2$P$_2$O$_7$. 

\begin{figure}[t]
\includegraphics[width=0.98\columnwidth]{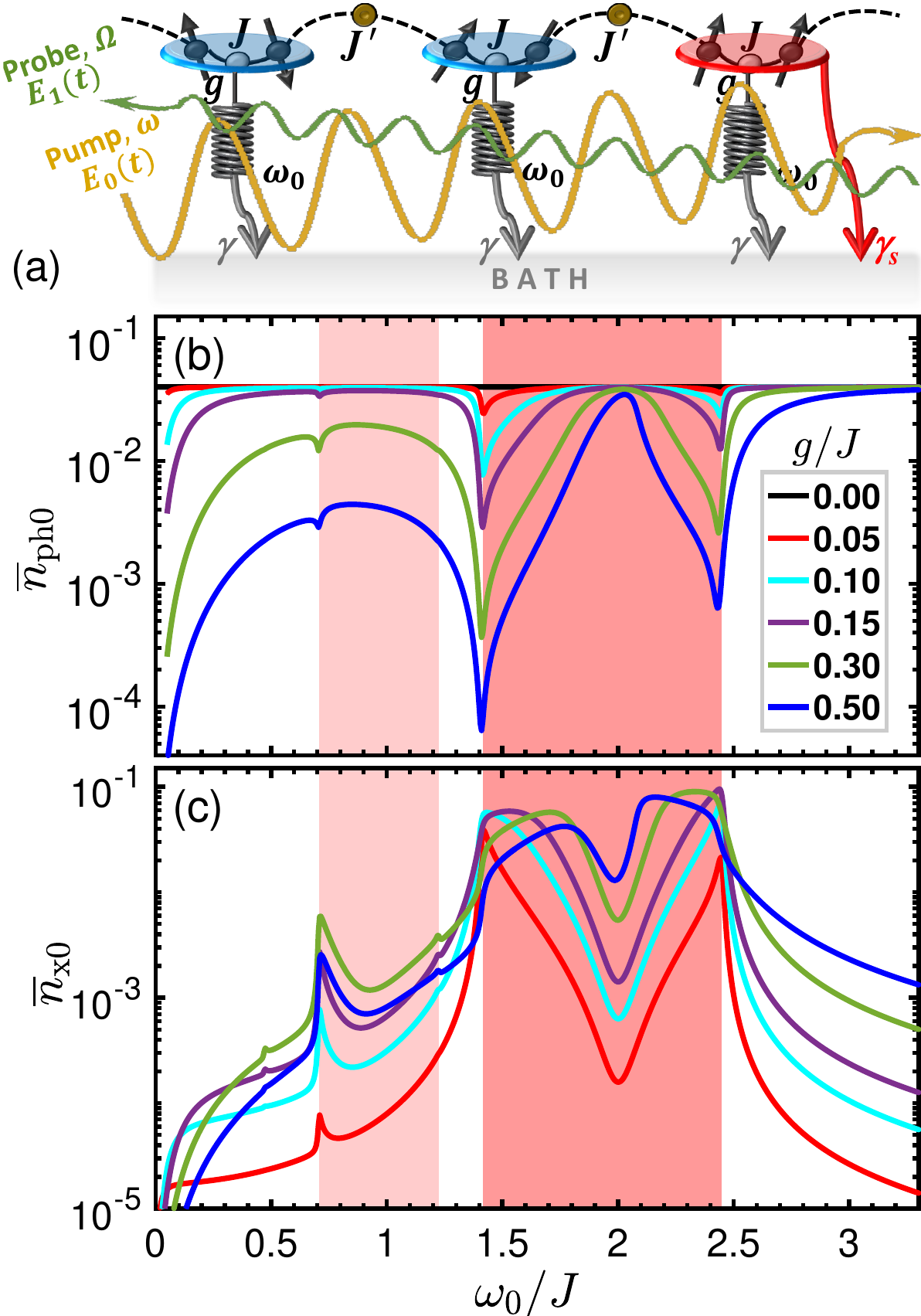}
\caption{{\bf Giant resonant self-blocking.}
(a) Schematic representation of the magnetophononically driven alternating 
spin chain with interaction parameters $J$ and $J'$, spin damping $\gamma_{\rm 
s}$, and spin-phonon coupling $g$; blue ellipses denote dimer singlets and 
the red ellipse a triplon excitation. The Einstein phonon frequency is 
$\omega_0$ and its damping is $\gamma$, the pump laser can drive the system at 
any frequency, $\omega$, and a probe beam addresses it at frequency $\Omega$. 
(b) Average driven phonon occupation, ${\overline n}_{\rm ph0} = n_{\rm ph0} 
(\omega = \omega_0)$, shown as a function of $\omega_0$ for $g/J = 0$, 0.05, 
0.1, 0.15, 0.3, and 0.5. The driving electric field is $E_0 = 0.2 \gamma$, 
with $\gamma = 0.02 \omega_0$ and $\gamma_s = 0.01 J$. Deep red shading marks 
the energy range, $2\omega_{\rm min} \le \omega_0 \le 2 \omega_{\rm max}$, of 
two-triplon excitations, light red shading the range where two-phonon harmonic 
processes create these excitations. (c) Corresponding average triplon 
occupation, ${\overline n}_{\rm x0} = n_{\rm x0} (\omega = \omega_0)$.}
\label{fgsb1}
\end{figure}


To analyze the dynamics of a phonon-driven and dissipative quantum magnet, we 
use the alternating $S = 1/2$ spin chain discussed in Ref.~\cite{rymfnu} and 
depicted in Fig.~\ref{fgsb1}(a). A bulk Einstein phonon, which is infrared 
(IR)-active to be excited coherently by the incident light, modulates one 
of the magnetic superexchange interactions at the driving frequency. The 
elementary excitations of the chain are gapped triplons, and the band of 
two-triplon modes excited by the driven phonon sets the resonant frequencies 
of the coupled system. Both this phonon and the spin excitations have a 
lifetime, i.e.~a damping, due to all the other (acoustic and optic) phonons 
of the lattice. 

We treat this open quantum system by introducing Lindblad operators 
\cite{rl}, the operators of the isolated system describing its interaction 
with the damping ``bath,'' to deduce the equations of motion governing 
the time evolution of its physical observables. A complete derivation may 
be found in Ref.~\cite{rymfnu} and, for a self-contained presentation, is 
summarized in Sec.~S1 of the Supplemental Material (SM) \cite{sm}. Here 
we state only our primary assumptions, that the Lindblad operators for 
the Einstein phonon are its own creation and annihilation operators, 
respectively $b_0^\dag$ and $b_0$, while those for the spin sector are the 
triplon creation and annihilation operators, ${\tilde t}_{k\alpha}^{\,\dag}$ 
and ${\tilde t}_{k\alpha}$, for each reciprocal-space mode of an $N$-dimer 
spin chain. These assumptions, whose physical meaning is also described 
in Sec.~S1, ensure a straightforward system of $3(N/2 + 1)$ equations of 
motion with no mixing between the triplon modes at different values of $k$. 


In the NESS established by steady laser driving, the phonon occupation, 
$n_{\rm ph} (t) = \big\langle {\textstyle \frac{1}{N}} b_0^\dag b_0 \big\rangle 
(t)$, oscillates at $2\omega$ about a finite average value, $n_{\rm ph0}$ 
\cite{rymfnu}. In Fig.~\ref{fgsb1}(b) we show $n_{\rm ph0}$ at $\omega = 
\omega_0$ (denoted ${\overline n}_{\rm ph0}$), with a laser electric-field 
strength ($E_0$ in energy units) and phonon damping ($\gamma$) that 
henceforth are held constant. At small $g$, ${\overline n}_{\rm ph0}$ 
is effectively constant for all $\omega_0$, but as $g$ is increased, 
${\overline n}_{\rm ph0}$ is suppressed precisely where the density of 
two-triplon excitations is highest. For $J'/J = 0.5$ as in Fig.~\ref{fgsb1}(b), 
the edges of the two-triplon band lie at $2 \omega_{\rm min} = 1.414 J$ and 
$2 \omega_{\rm max} = 2.449 J$ \cite{rymfnu}, and this resonant effect becomes 
gigantic at strong $g$, suppressing the phonon occupation by nearly three 
orders of magnitude at $2 \omega_{\rm min}$.

\begin{figure*}[t]
\includegraphics[width=0.93\textwidth]{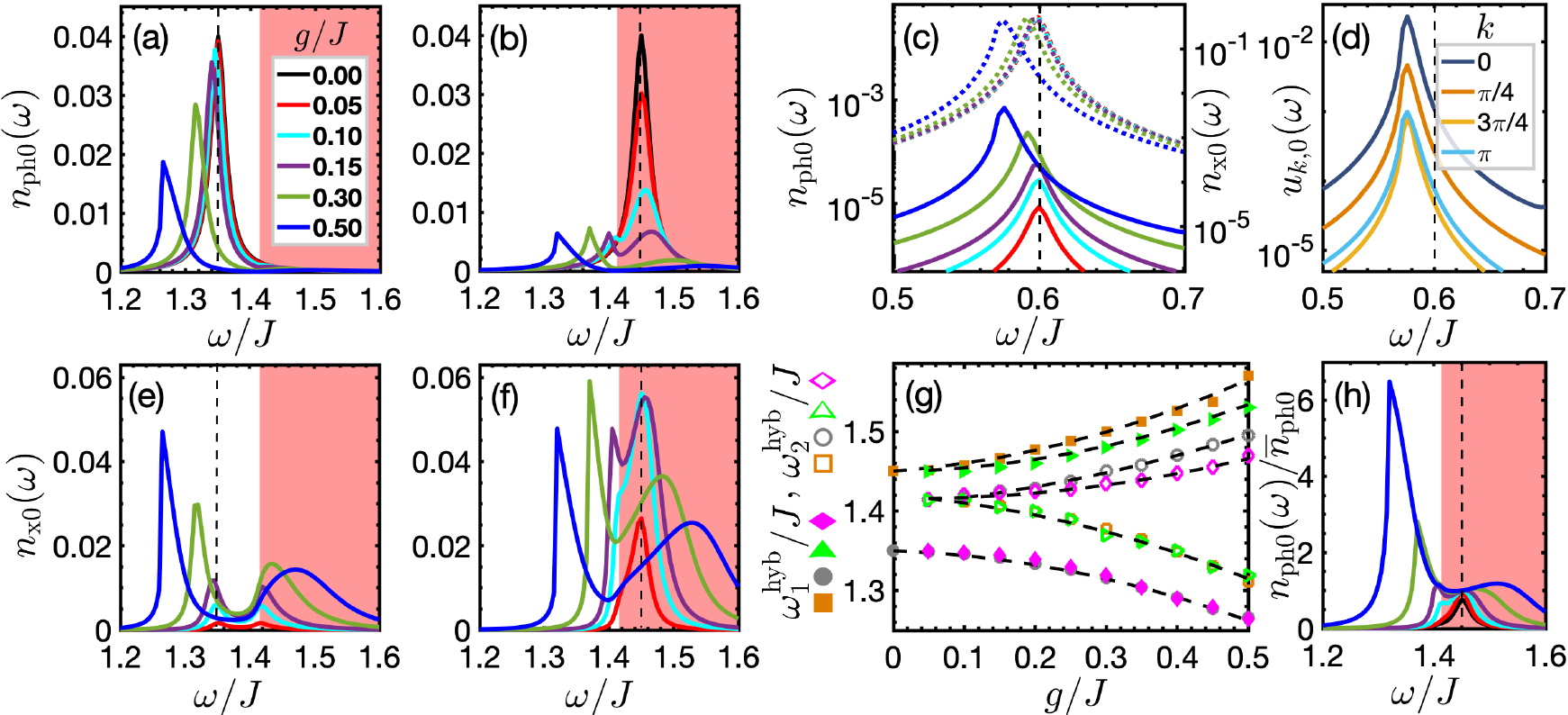}
\caption{{\bf Strongly and weakly hybridized excitations.} 
(a) Phonon occupation, $n_{\rm ph0}$, shown as a function of driving frequency, 
$\omega$, for a phonon frequency $\omega_0 = 1.35 J$ at selected $g$ values. 
The standard driving and damping parameters of Fig.~\ref{fgsb1}(b) are used. 
(b) As in panel (a) for $\omega_0 = 1.45 J$. 
(c) $n_{\rm ph0} (\omega)$ (dotted lines) and triplon occupation, $n_{\rm x0} 
(\omega)$ (solid lines), shown for a phonon frequency, $\omega_0 = 0.6 J$, 
far from the spin excitation band. 
(d) $k$-resolved components of the average $u_k (\omega)$ at $\omega_0 = 0.6 J$.
(e) $n_{\rm x0} (\omega)$ for $\omega_0 = 1.35 J$, corresponding to panel (a). 
(f) $n_{\rm x0} (\omega)$ for $\omega_0 = 1.45 J$, corresponding to panel (b). 
(g) Peak-pair frequencies, labelled $\omega_1^{\rm hyb}$ and $\omega_2^{\rm hyb}$, 
as in panels (a), (b), (e), and (f), shown for all $g$ values; black dashed 
lines indicate a $g^2$ form. 
(h) Data of panel (b) normalized to the resonant phonon occupation, ${\overline 
n}_{\rm ph0}$, at the same $\omega$ [data of Fig.~\ref{fgsb1}(b)].}
\label{fgsb2}
\end{figure*}

We have named this effect ``self-blocking'' because the magnetic system acts 
to block its own energy uptake by blocking the driven phonon. This behavior 
is surprising if one expects stronger energy absorption when more spin 
excitations coincide with the driving laser frequency. Its explanation lies 
in the fact \cite{rymfnu} that in magnetophononic driving the spin system 
is not coupled to the light, but only to the driven phonon. Heuristically, it 
acts as an extra ``inertia'' for the phonon to drive into motion. Analytically, 
the prefactor of the phonon momentum, $p(t) = \big\langle {\textstyle \frac{i}
{\sqrt{N}}} (b_0^\dag - b_0) \big\rangle (t)$, in the master equation for 
$n_{\rm ph} (t)$ contains terms by which the spin system acts directly against 
the driving electric field, suppressing its effective value to ${\widetilde 
E}_0 (t) = E_0(t) - g \sum_k \langle f({\tilde t}_{k\alpha}^{\,\dag}, {\tilde 
t}_{k\alpha}) \rangle$ (Secs.~S1 and S2 of the SM \cite{sm}). This negative 
feedback effect is strongly nonlinear in $g$ and can cancel $E_0 (t)$ almost 
completely at resonance [Fig.~\ref{fgsb1}(b)]. Despite the approximate 
symmetry of the spin band, self-blocking is weaker by a factor of 10 at 
$2 \omega_{\rm max}$ due to matrix elements within the feedback process.

Away from the two-triplon band, in Fig.~\ref{fgsb1}(b) we observe a 
significant suppression of phonon energy entering the system at any 
frequency $\omega_0 < 2 \omega_{\rm min}$. This nonresonant self-blocking 
is also nonlinear in $g$, exceeding one order of magnitude at $g = 0.5 J$. 
Its appearance only in the low-$\omega_0$ regime, but not at $\omega_0 > 2 
\omega_{\rm max}$, points to an origin in multiple harmonic processes 
($2\omega_{\rm min} \le n \omega_0 \le 2 \omega_{\rm max)}$ \cite{rymfnu}. 
Although only the two-phonon harmonic ($n = 2$) at $\omega_{\rm min}$ is 
visible directly, stronger $g$ distributes the response of the system to 
a given $n \omega_0$ across a broader range of frequencies. By contrast, 
a driving phonon at the band center ($\omega_0 = 2J$) has vanishing matrix 
elements with the resonant spin modes, and hence ${\overline n}_{\rm ph0}$ 
recovers almost to its $g = 0$ value for all $g$. 

Turning to the response of the spin system, Fig.~\ref{fgsb1}(c) shows the 
corresponding average triplon occupancy, ${\overline n}_{\rm x0}$. The most 
striking feature is the strong rounding of the in-band response as $g$ is 
increased. The band-edge peaks are entirely blunted by the strong suppression 
of ${\overline n}_{\rm ph0}$ [Fig.~\ref{fgsb1}(b)]. We stress that the effective 
limiting value ${\overline n}_{\rm x0} \approx 0.1$ visible in Fig.~\ref{fgsb1}(c)
is purely a consequence of the giant self-blocking, and is not connected with 
the hard-core nature of the triplon excitations, which has not been included 
in our formalism \cite{rymfnu}. This rounding suggests an increasing 
localization of the spin response, by which the band character of the triplons 
becomes less relevant under strong driving by the entirely local phonon. 


In Figs.~\ref{fgsb1}(b) and \ref{fgsb1}(c) the system is driven at the phonon 
frequency, $\omega_0$. It is important to note that self-blocking is not a 
simple shift of phonon energy to different frequencies: NESS established by 
steady driving contain no frequencies other than those of the drive and its 
higher harmonics \cite{rymfnu}. To probe how a strong $g$ modifies the spin 
and phonon spectra, we begin with the phononic response of a system driven 
at frequencies $\omega \ne \omega_0$. Focusing first on $\omega_0$ values near 
resonance, Fig.~\ref{fgsb2}(a) shows a phononic excitation that lies just 
below the two-triplon band being weakened and pushed away from the lower band 
edge at stronger $g$, i.e.~a level repulsion. Figure \ref{fgsb2}(b) shows the 
analogous result when $\omega_0$ lies just inside the spin band, where the 
phonon peak is damped very strongly with increasing $g$, and is also repelled 
from the band edge. Here it is accompanied by the development of a second 
feature, appearing at $2\omega_{\rm min}$ at $g = 0.1 J$, which is repelled 
below the band edge as $g$ increases. 

Before proceeding, the giant self-blocking at resonance raises the question 
of whether this two-peak effect could be just a secondary consequence of the 
very strongly suppressed phononic response at $\omega = 2\omega_{\rm min}$.
For a heuristic measure of self-blocking we show in Fig.~\ref{fgsb2}(h) the 
result of Fig.~\ref{fgsb2}(b) normalized by ${\overline n}_{\rm ph0}$ from 
Fig.~\ref{fgsb1}(b). The weak-$g$ peaks then appear with unit magnitude, 
while the strong-$g$ phononic response does confirm a two-peak structure 
with level repulsion, indicating the formation of hybrid spin-phonon states. 
Here the in-band hybrid remains damped, whereas the hybrid generated outside 
the spin band responds much more strongly per unit driving. 


To confirm the hybrid nature of these states we examine their spin character. 
When $\omega_0$ is far from the two-triplon band, $n_{\rm x0} (\omega)$ indicates 
that a magnetic response emerges with $g$ despite the nonresonant self-blocking 
[Fig.~\ref{fgsb2}(c)]. The minor changes in $n_{\rm ph0} (\omega)$ indicate that 
this is a localized phononic mode whose weak hybridization is enough to shift 
its frequency out of Fig.~\ref{fgsb1}(c) and whose dressing at $g = 0.5 J$ 
involves all the $k$-components of $n_{\rm x0} (\omega) = \frac{1}{N} \sum_k 
u_{k,0} (\omega)$ ($u_k = \sum_\alpha \langle {\tilde t}_{k\alpha}^{\,\dag} {\tilde 
t}_{k\alpha} \rangle$) [Fig.~\ref{fgsb2}(d)]. Returning to the band edges, 
Figs.~\ref{fgsb2}(e) and \ref{fgsb2}(f) complement Figs.~\ref{fgsb2}(a) and 
\ref{fgsb2}(b), and in Fig.~\ref{fgsb2}(g) we gather the characteristic 
frequencies of these phonon and spin spectra [including the second peak 
weakly visible inside the band in Fig.~\ref{fgsb2}(a), highlighted on a 
logarithmic scale in Sec.~S2 of the SM \cite{sm}]. These display the clear 
development with $g$ of two mutually repelling hybrid excitations whose 
frequency shifts scale accurately with $g^2$; we show in Sec.~S3 that the 
same physics is also found for driving frequencies around the upper band edge. 

Concerning the admixture of lattice and spin character, if one defines a 
hybridization parameter $s = g/|\omega_0 - 2 \omega_{\rm min}|$, then a language 
of ``phononic'' and ``magnetic'' hybrids remains useful at $\omega_0 = 0.6 J$ 
[Fig.~\ref{fgsb2}(c)], where $s < 1$ for all $g$. However, when $s \approx 10$ 
both hybrids are strongly magnetic and phononic, and indeed the 50:50 weight 
distribution at larger $g$ in Figs.~\ref{fgsb2}(e) and \ref{fgsb2}(f) suggests 
states that are maximally hybridized. For the hybrids repelled outside the 
band, the coinciding peaks in $n_{\rm ph0} (\omega)$ and $n_{\rm x0} (\omega)$ 
identify them as a strongly triplon-dressed version of the ``phononic'' hybrid 
shown in Figs.~\ref{fgsb2}(c) and \ref{fgsb2}(d). The in-band hybrids lie in a 
continuum of propagating triplon-pair states, and thus manifest themselves as 
broader peaks, lying at slightly different energies, in $n_{\rm ph0} (\omega)$ 
[Fig.~\ref{fgsb2}(b)] and $n_{\rm x0} (\omega)$ [Fig.~\ref{fgsb2}(f)].

For the driving and damping of our system, all hybrid states are to a good 
approximation ``phonon-bitriplons,'' i.e.~phonons dressed by one triplon pair 
(${\tilde t}_k^\dag {\tilde t}_{-k}^{\,\dag}$) of zero net momentum. Despite the 
ubiquity in physics of ``light-matter'' interaction processes where a boson 
produces pairs of fermions, processes where one boson produces pairs of 
bosons are rare. Phonon-bimagnon processes have been discussed both 
theoretically \cite{rlsl,rlsb} and experimentally \cite{Grueninger2000,
Windt2001} in the optical spectra of cuprate quantum magnets. Similar physics 
could be engineered using ultracold atoms \cite{rec}, where the optical lattice 
blurs the distinction between photon and phonon, although we are not aware of 
an experiment. On a very different energy scale, in particle physics the 
virtual decay of the Higgs boson into pairs of W or Z bosons \cite{rhdwz,rsea} 
is an off-shell process with intermediate $s$, where the level repulsion of 
Fig.~\ref{fgsb2}(g) is known as a ``Higgs mass renormalization.'' 

\begin{figure}[t]
\includegraphics[width=\columnwidth]{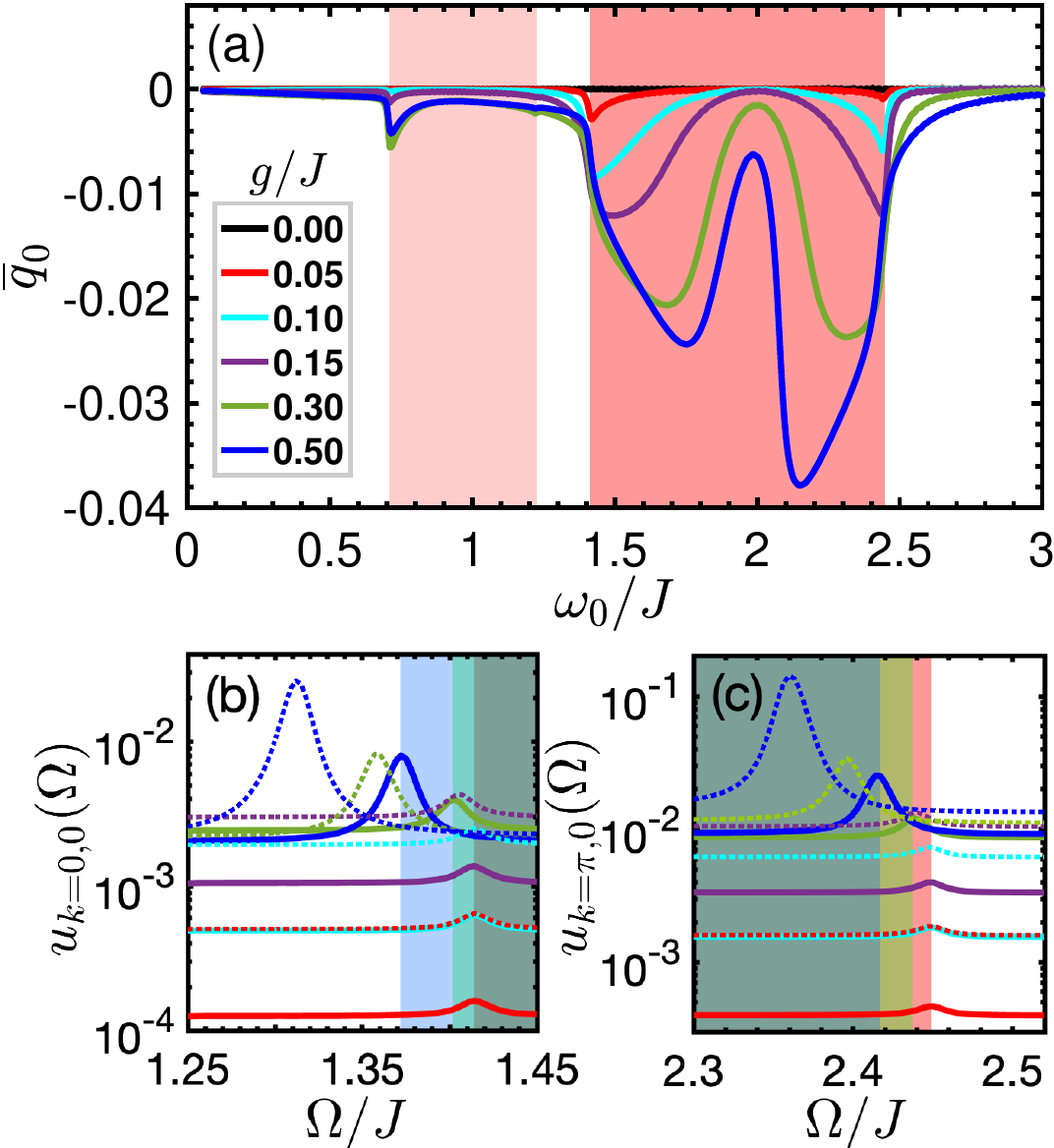}
\caption{{\bf Dynamical spin-phonon renormalization.}
(a) Static displacement of the Einstein phonon, ${\overline q}_0$, resulting 
from standard driving at frequency $\omega = \omega_0$, shown for a selection 
of $g$ values. (b) Average value of $u_{k=0}$, shown as a function of the probe 
frequency, $\Omega$, for a standard driving field $E_0$ at $\omega = \omega_0
 = 2.2 J$ (solid lines) and for driving with $2E_0$ (dotted lines); 
the probe field is set to $E_1 = 0.2E_0$. (c) As in panel (b) for $u_{k=\pi}$. 
Blue, green, red, and superposed shading indicates the corresponding 
driving-renormalized two-triplon bands.} 
\label{fgsb3}
\end{figure}


Having demonstrated both strong driving-induced changes to the excitation 
spectra and strong suppression of this driving near resonance, we consider 
the consequences for the properties of the NESS. Figure \ref{fgsb3}(a) 
shows that ${\overline q}_0$, the average of the phonon displacement [$q(t) = 
\big\langle {\textstyle \frac{1}{\sqrt{N}}} (b_0^\dag + b_0) \big\rangle (t)$] 
obtained when driving at $\omega = \omega_0$, is pushed to a finite value, 
which can reach 4\% of the lattice dimension at $g = 0.5 J$. In our minimal 
model, this driven distortion implies a linear modification of the 
magnetic interaction $J$ to ${\tilde J} = J + g {\overline q}_0$. The 
associated renormalization of the two-triplon band is partly reflected in 
the evolution of $u_{k0} (\omega)$ on increasing $g$, as we show in Sec.~S4 
of the SM \cite{sm}. However, for an accurate measurement we introduce the 
``pump-probe'' protocol of Fig.~\ref{fgsb1}(a), driving the system at $\omega
 = \omega_0$ and monitoring its response to an additional, weak probe 
component of the laser at frequency $\Omega$. Because we are investigating 
NESS, $E(t) = E_0 \cos (\omega t) + E_1 \cos (\Omega t)$ is continuous 
and the time delay used in true pump-probe studies is absent. 

As the most sensitive diagnostic of the edges of the modified two-triplon 
band, in Figs.~\ref{fgsb3}(b) and \ref{fgsb3}(c) we show not $n_{\rm x0} 
(\omega_0,\Omega)$ but the respective components $u_{k=0,0}$ and $u_{k=\pi,0}$. 
As $\Omega$ is scanned through the band-edge frequencies, the peaks in both 
quantities grow and shift away from the equilibrium band edges when $g = 0.3 
J$ and 0.5$J$. In contrast to Figs.~\ref{fgsb2}(e) and \ref{fgsb2}(f), where 
two strongly hybridized excitations form at the band edge due to the 
near-resonant phonon mode, $s < 1$ when the system is driven at $\omega = 
\omega_0 = 2.2 J$ and we observe a single peak in the magnetic response. When 
corrected for the corresponding weak hybridization shift, $\delta \omega_s < 
0.005 J$, which we show in Sec.~S4 of the SM \cite{sm}, these peaks indicate 
a driving-induced renormalization of the entire two-triplon band. Assuming 
that only ${\tilde J}$ is renormalized yields quantitative agreement with the 
peaks observed in our pump-probe spectra, i.e.~the triplon band retains its 
cosine form (also shown in Sec.~S4) in the model of Fig.~\ref{fgsb1}(a).

This dynamical renormalization is again a strongly nonlinear function of $g$, 
and demonstrates how magnetophononic driving can be used to control the spin 
states at frequencies far from that of the pump. The key physics contained in 
Fig.~\ref{fgsb3}(a) is that the phonon frequencies most effective for exerting 
this control are neither those at the band edges, where giant self-blocking 
[Fig.~\ref{fgsb1}(b)] suppresses ${\overline q}_0$ almost completely, nor those 
at the band center, where the driving terms decouple, but the ``quarter-band'' 
ones around $k = \pi/4$ and $3\pi/4$. Considering a model with the phonon 
coupled instead to the $J'$ bond in Fig.~\ref{fgsb1}(a) reinforces these 
conclusions while allowing a different type of band renormalization, 
as we show in Sec.~S5 of the SM \cite{sm}. Although the band shifts in 
Figs.~\ref{fgsb3}(b) and \ref{fgsb3}(c) are not large for our standard 
driving (solid lines), doubling the electric field leads to a very strong 
effect (dotted lines), such that a majority of the total weight in the 
nonequilibrium (driven) spin spectrum can appear at frequencies that 
are forbidden at equilibrium. 


Turning to experiment, CuGeO$_3$ \cite{rwgb} and (VO)$_2$P$_2$O$_7$ \cite{run2} 
are quantum magnetic materials known to have very large $g$. Both have low 
structural symmetry, making IR-active phonons available over a range of 
energies around relatively broad spin bands \cite{Popovic95,Regnault96,
Uhrig97,Grove00,run2}. Experiments to create spin NESS at the resonant 
frequencies of bulk-driven quantum magnets require a thin-film geometry and 
efficient thermal transfer to maintain a low sample temperature \cite{rymfnu}. 
In principle, self-blocking can allow significant relaxation of the constraints 
on pump intensity, driving time, and sample thickness, although in practice 
strong electromagnetic driving can induce heating by a variety of 
channels. For controlling the spin excitation spectrum, we comment that 
${\tilde J} (q_0)$ is in general a highly nonlinear function, and although
conventional experimental probes usually require only perturbative expansions, 
coherent laser driving can produce very large $q_0$ values \cite{Giorgianni21}.


Self-blocking is favored by a high density of spin states. Thus it should 
be prevalent not only at the band edges in low-dimensional quantum magnets 
but also in any system with nearly flat spin bands, which can arise in strongly 
frustrated materials of any dimensionality. SrCu$_2$(BO$_3$)$_2$ is one such 
system \cite{Miyahara03}, in which a recent experiment has demonstrated a 
phonon-driven nonequilibrium population of two-triplon excitations
\cite{Giorgianni21}, but not yet a self-blocking. We stress also that the 
dissipative processes in our analysis are generic and hence the phenomenology 
of self-blocking is independent of the nature of the bath, as discussed in 
Sec.~S1 of the SM \cite{sm}. 

Coherent phononic driving can be applied in quantum magnetic materials 
at high phonon frequencies to Floquet-engineer collective spin states and 
with slow phonons to modulate the existing magnetic energy levels. Resonant 
magnetophononic driving goes beyond these situations by creating qualitatively 
different hybrid quantum states and feedback effects. In this regime we have 
discovered an initially counterintuitive giant self-blocking. We explain 
this phenomenon and place it in the context of controlling the dynamical 
renormalization of spin states by light through the medium of the lattice. 

\begin{acknowledgments}
{\it Acknowledgments.} We thank F.~B.~Anders, D.~Bossini, B.~Fauseweh, 
F.~Giorgianni, C.~Meyer, Ch.~R\"uegg, and L.~Spitz for helpful discussions. 
Research at TU Dortmund University was supported by the German Research 
Foundation (DFG) through Grant No.~UH 90-13/1 and together with the Russian 
Foundation of Basic Research through project TRR 160. 
\end{acknowledgments}


\setcounter{equation}{0}
\renewcommand{\theequation}{S\arabic{equation}}
\setcounter{figure}{0}
\renewcommand{\thefigure}{S\arabic{figure}}
\setcounter{section}{0}
\renewcommand{\thesection}{S\arabic{section}}
\setcounter{table}{0}
\renewcommand{\thetable}{S\arabic{table}}

\onecolumngrid

\vskip12mm

\centerline{\large {\bf {Supplemental Material for ``Giant Resonant 
Self-Blocking}}} 

\vskip1mm

\centerline{\large {\bf {in Magnetophononically Driven Quantum Magnets''}}}

\vskip4mm

\centerline{M. Yarmohammadi, M. Krebs, G. S. Uhrig, and B. Normand}

\vskip8mm

\twocolumngrid

\section{S1.  Quantum Master Equations}

\subsection{A.  Coupled phonon and spin systems}

We consider a minimal Hamiltonian describing a straightforward and well 
understood quantum spin system coupled strongly to a single, nondispersive 
optical phonon. A detailed derivation of the appropriate diagonal Hamiltonians 
and the equations of motion governing the physical observables is presented in 
Ref.~\cite{rymfnu}. This subsection provides a summary of the methods, 
notation, and underlying physics. 

The quantum spin system is an alternating $S = 1/2$ Heisenberg chain, which 
can be described in terms of the operators $\tilde{t}_{k\alpha}^{\,\dagger}$ and 
$\tilde{t}_{k\alpha}$ creating and destroying triplon excitations [Fig.~1(a) 
of the main text]. The spin Hamiltonian in reciprocal space takes the form 
\begin{equation} 
H_\text{s} = \sum_{k, \alpha} \omega_k \tilde{t}_{k\alpha}^{\,\dag} \tilde{t}_{k\alpha},
\label{ehsd}
\end{equation} 
with the dispersion relation 
\begin{equation} 
\omega_k = J \sqrt{1 - \lambda \cos k}, 
\label{ewk}
\end{equation} 
where $\lambda = J'/J$. 

For illustration we consider an Einstein phonon mode coupling strongly to 
only one of the magnetic interactions of the spin system and driven by the 
electric field of the laser. The relevant Hamiltonian terms in real space are
\begin{eqnarray}
H_\text{p} + H_\text{sp} + H_\text{l} & = & 
\sum_j  \big[ \omega_0 b_j^\dag b_j + g (b_j + b_j^\dag) 
\, {\vec S}_{1,j} \! \cdot \! {\vec S}_{2,j} \nonumber \\ 
& & \qquad\qquad\quad + \, E (t) (b_j + b_j^\dag) \big],
\label{ehp}
\end{eqnarray}
where ${\vec S}_{1,j}$ and ${\vec S}_{2,j}$ are the $S = 1/2$ spin operators on 
dimer $j$ of the chain, $g$ is the spin-phonon coupling constant, and $E (t)
 = E_0 \cos (\omega t)$ is the single-frequency oscillating electric field of 
the light. To describe the physical observables of the driven phonon we 
consider the real variables 
\begin{subequations}
\label{eq:expval-phon}
\begin{eqnarray}
q(t) & = & \big\langle {\textstyle \frac{1}{\sqrt{N}}} (b_0 + b_0^\dag) 
\big\rangle (t), \\ 
p(t) & = & \big\langle {\textstyle \frac{i}{\sqrt{N}}} (b_0^\dag - b_0) 
\big\rangle (t), \\  
n_{\rm ph}(t) & = & \big\langle {\textstyle \frac{1}{N}} b_0^\dag b_0 \big\rangle 
(t), 
\label{erpv}
\end{eqnarray}
\end{subequations}
at any given time, $t$, which correspond respectively to the displacement of 
the Einstein phonon, the conjugate momentum, and the number operator. 

To formulate analogous quantities for the spin sector we consider the 
$k$-space components of the triplon number operator and its off-diagonal 
equivalent,
\begin{subequations}
\label{ersv}
\begin{eqnarray}
u_k & = & \sum_\alpha  \tilde{t}_{k\alpha}^{\,\dag} \tilde{t}_{k\alpha} \;\; 
{\rm and} \\ {\tilde{v}}_k & = & \sum_\alpha \tilde{t}_{k\alpha}^{\,\dag} 
\tilde{t}_{-k\alpha}^{\,\dag}. 
\end{eqnarray}
\end{subequations}
We denote their expectation values by 
\begin{subequations}
\label{eq:expect}
\begin{eqnarray}
\label{eq:u-def}
u_k(t) & = & \langle u_k \rangle (t) \\ 
\tilde v_k(t) & = & \langle {\tilde{v}}_k \rangle (t),
\end{eqnarray}
\end{subequations}
and similarly for $\tilde v_k^*(t)$, then separate the real and imaginary 
parts to obtain
\begin{subequations}
\label{eq:vw-def}
\begin{align}
v_k(t) & = {\rm Re} \, \tilde v_k(t) \\
w_k(t) & = {\rm Im} \, \tilde v_k(t).
\end{align}
\end{subequations}
$u_k(t)$ is a real variable and a valuable characterization of the driven 
spin sector is provided by the nonequilibrium triplon number, 
\begin{equation}
n_\text{x}(t) = \frac{1}{N} \sum_k u_k(t).
\label{enx}
\end{equation}

It is convenient to define the coefficients 
\begin{subequations}
\begin{eqnarray}
y_k & = & \frac{1 - {\textstyle \frac12} \lambda \cos k}{\sqrt{1 - \lambda 
\cos k}} = \frac{J}{2} \frac{1 + \omega^2_k / J^2}{\omega_k} \;\;\; {\rm and} 
\label{eyk} \\ y'_k & = & \frac{ {\textstyle \frac12} \lambda \cos k}{\sqrt{1
 - \lambda \cos k}} = \frac{J}{2} \frac{1 - \omega^2_k / J^2}{\omega_k}, 
\label{eypk}
\end{eqnarray}
\end{subequations}
and the real quantities
\begin{subequations}
\label{ecuv}
\begin{align}
\mathcal{U}(t) & = \frac{1}{N} \sum_k y_k [u_k(t) - 3 n(\omega_k)] \;\;\; {\rm 
and} \\ \mathcal{V}(t) & = \frac{1}{N} \sum_k y_k' v_k(t),
\end{align}
\end{subequations}
where ${n}(\omega_k) = [\exp (\hbar\omega_k / k_{\rm B} T) - 1]^{-1}$, the 
bosonic occupation function for the triplon mode with frequency $\omega_k$, 
is an accurate approximation to the true occupancy of the hard-core triplon 
modes for small $n_{\rm ph}$. 

\subsection{B.  Equations of motion}

The time evolution of an open quantum system is specified by adjoint 
quantum master equations \cite{rbp} of the form 
\begin{eqnarray}
\label{eaqme}
&& \frac{d}{dt} A_\text{H}(t) = i [H, A_\text{H}(t)] \\ && \quad + \sum_{l} 
\tilde\gamma_l \big[ A_l^\dag A_\text{H}(t) A_l - {\textstyle \frac12} 
A_\text{H}(t) A_l^\dag A_l - {\textstyle \frac12} A_l^\dag A_l A_\text{H}(t) \big] 
\nonumber
\label{eq:lindblad1}
\end{eqnarray}
for any operator $A_\text{H}(t)$ describing a physical observable. In these 
Heisenberg equations of motion, $H$ is the Hamiltonian of the isolated 
quantum system, excluding the bath. The Lindblad operators, $\{A_l\}$, 
are also formed from those of the isolated system and the coefficients 
$\tilde\gamma_l$ are effective damping parameters. The Lindblad operators 
for the driven phonon are $A_1 = b_0^\dag$ and $A_2 = b_0$, with corresponding 
damping coefficients $\gamma_1 = \gamma n(\omega_0)$ and $\gamma_2 = \gamma 
[1 + n(\omega_0)]$, which return the equations of motion of the damped 
harmonic oscillator,
\begin{subequations}
\label{eq:eom-phonon}
\begin{eqnarray}
\frac{d}{dt} q(t) & = & \omega_0 p(t) - {\textstyle {\frac12}} \gamma 
q(t), \label{eompa} \\
\frac{d}{dt} p(t) & = & - \omega_0 q(t) - {\textstyle {\frac12}} \gamma p(t) 
 - 2 {\widetilde E}(t), 
\label{eompb} \\
\frac{d}{dt} n_{\rm ph}(t) & = & - {\widetilde E}(t) p(t) - \gamma [n_{\rm ph}(t)
 - {n}(\omega_0)], 
\label{eompc}
\end{eqnarray}
\end{subequations}
in which ${\widetilde E}(t) = E(t) + g (\mathcal{U}(t) + \mathcal{V}(t))$ 
expresses the effective electric field of the light acting on the Einstein 
phonon in the presence of renormalization arising from the spin system to 
which the phonon is coupled. 

Turning to the spin system, we adopt as the Lindblad operators the linear 
one-triplon operators, $\tilde{t}_k$ and $\tilde{t}_k^\dag$, with a single 
spin-damping coefficient, $\tilde \gamma_{k} = \gamma_\text{s} n(\omega_k)$. 
As discussed in Ref.~\cite{rymfnu}, these are not spin-conserving operators: 
physically, their meaning is that a phonon oscillation can change the spin 
quantum number and thus they are appropriate for systems with appreciable 
spin-orbit coupling. In a real material, their effects would be accompanied 
by those of bilinear Lindblad operators (of the form $C_{kq} = {\tilde t}_k^\dag 
{\tilde t}_q$), which would be the leading bath terms in a system with weak 
spin anisotropy. However, here we use only the linear operators for simplicity 
of presentation and for physical transparency. The equations of motion of the 
spin sector are then 
\begin{subequations}
\label{eq:eoms}
\begin{eqnarray} 
\!\!\!\!\!\! \frac{d}{dt} u_k(t) & = & 2g q (t) y'_k  w_k (t) \! - \! 
\gamma_\text{s} [u_k (t) \! - \! 3 n(\omega_k) ], \label{eomsu} \\
\!\!\!\!\!\! \frac{d}{dt} v_k (t) & = & -2 [\omega_k + g y_k q(t)] w_k (t) 
- \gamma_\text{s} v_k (t), \label{eomsv} \\ 
\!\!\!\!\!\! \frac{d}{dt} w_k (t) & = & 2 [\omega_k + g y_k q(t)] v_k (t) 
\\ & & + 2g q (t) y'_k \left[ u_k(t) + {\textstyle \frac{3}{2}} 
\right] - \gamma_\text{s} w_k (t), \nonumber 
\label{eomsw}
\end{eqnarray}
\end{subequations}
for each mode $k$ of the spin chain. 

\begin{figure}[t]
\includegraphics[width=\columnwidth]{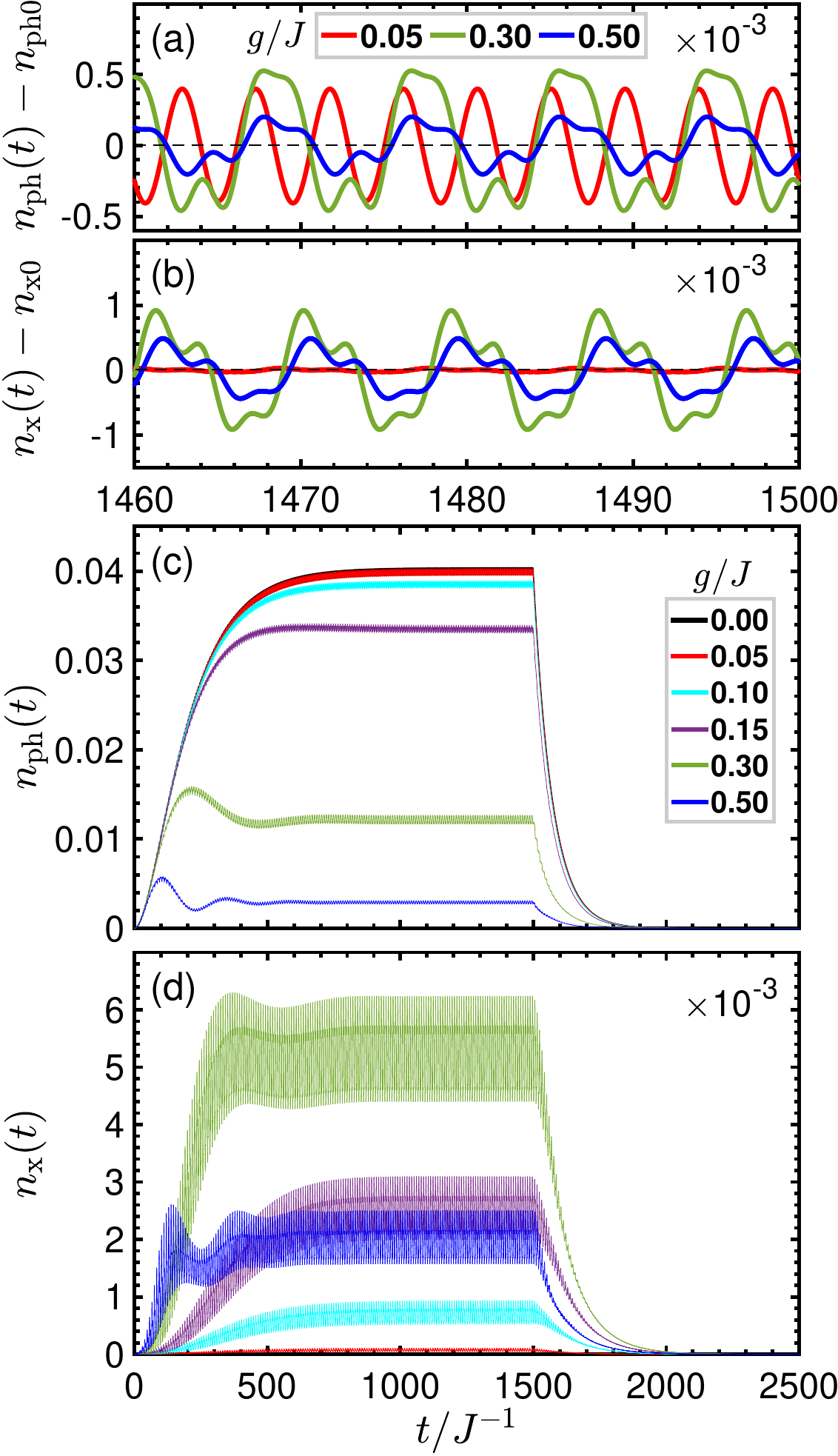}
\caption{Example of phonon (a) and spin NESS (b) shown as a function of time 
for typical driving and damping parameters. The phonon frequency, $\omega_0 
 = 0.707 J$, is set to half the lower band-edge frequency of the isolated 
spin system, a value at which it shows some weak higher harmonic effects 
[Figs.~1(b) and 1(c) of the main text]. Development of the phonon (c) and 
spin response (d) from $t = 0$, illustrating how feedback effects from the 
spin sector arrest the growth of the phonon occupation at a very early stage, 
although the slow, damped oscillations in the average values mean that the 
NESS is reached at approximately the same time in all cases.}
\label{fgsbsm1}
\end{figure}

We solve the $3(N/2+1)$ equations of motion for spin chains with lengths up 
to $N = 2000$ dimers. We focus on the nonequilibrium steady states (NESS) of 
the driven, dissipative system, whose formation at small $g$ values requires a 
timescale of approximately $5/\gamma_{\rm s}$, i.e.~five time constants of the 
spin system \cite{rymfnu}. However, NESS formation at large $g$ values may 
require several cycles through strong feedback processes, and to deal with  
these cases we consider times up to $t = 30\,000/J$. An example of 
complementary phonon and spin NESS, each characterized by their number 
operators, is shown in the time domain for a nonresonant system frequency, 
$\omega_0 = \omega_{\rm min}$, in Fig.~\ref{fgsbsm1} at weak, strong, and very 
strong values of $g$. We observe that both time traces contain increasingly 
complex combinations of harmonics as $g$ is increased; the amplitude of the 
oscillatory part of the phonon occupation starts to be suppressed at very 
strong $g$; the oscillatory part of the triplon occupation rises very strongly 
with $g$ on exiting the weak-coupling regime, but is also suppressed at very 
strong coupling. The corresponding static parts of the both occupations, 
${\overline n}_{\rm ph0}$ and ${\overline n}_{\rm x0}$, may be read respectively 
from Figs.~1(b) and 1(c) of the main text for all phonon frequencies.

Concerning the timescale for the development of self-blocking, 
Fig.~\ref{fgsbsm1}(c) shows how the initial rise of $n_{\rm ph}$ is truncated 
by the rise of $n_{\rm x}$ [Fig.~\ref{fgsbsm1}(d)]. In this nonresonant regime, 
at $g = 0.3 J$ there remains a significant time lag between the driving phonon 
and following triplon occupations, where the latter limits the former and 
convergence requires one cycle. At $g = 0.5 J$, the lag in response is much 
shorter and several slow oscillation cycles are required. The same phenomena 
are visible at $\omega_0 = 2 \omega_{\rm min}$ (not shown), except that the 
resonant self-blocking is so extreme that it is difficult to see the 
$n_{\rm ph}(t)$ curves for all $g$ on the same scale, and convergence can 
require more than 5 slow oscillation cycles.

We comment that the equations of motion are valid at all times from the onset 
of driving ($t = 0$) to infinity and for all applied electric fields, as well 
as for all phonon occupations up to the Lindemann melting criterion ($n_{\rm ph} 
\approx 3$). With the present simplified treatment of the spin sector, they 
are valid up to a triplon occupation of order $n_{\rm x} \approx 0.2$, beyond 
which a more sophisticated numerical treatment should be used to account for 
the hard-core occupation constraint. Because the equations of motion are based 
on a mean-field decoupling of the spin and lattice sectors, our treatment 
becomes more approximate at low phonon frequencies \cite{rymfnu}, specifically 
those below $\omega_0 = 0.2$--$0.3J$ on the left side of Fig.~1 of the main 
text. Nevertheless, one may verify by considering the energy flow through the 
strongly spin-phonon-coupled system that the mean-field approximation remains 
very accurate at resonance ($\omega_0 = 2 \omega_{\rm min}$), i.e.~that its 
deterioration is a consequence of the frequency range, not only of a very 
low energy flux within the system, and thus that it is not a factor in 
self-blocking. 

\begin{figure}[t]
\includegraphics[width=\columnwidth]{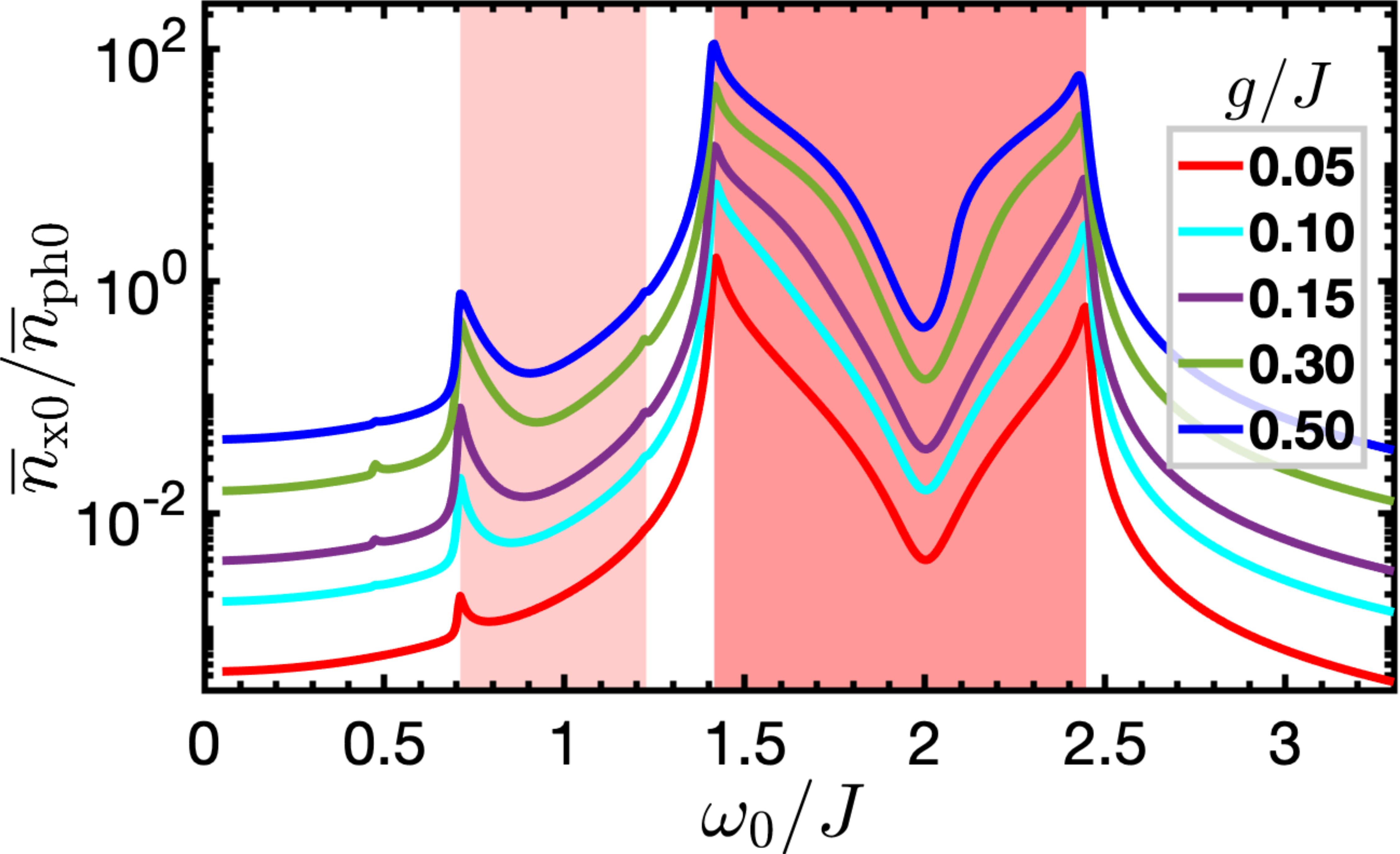}
\caption{Driven triplon occupation, ${\overline n}_{\rm x0}$ from Fig.~1(c) 
of the main text, normalized by the actual phonon occupation, ${\overline 
n}_{\rm ph0}$, taken from Fig.~1(b) of the main text.}
\label{fgsbsm2}
\end{figure}

Finally, one may question the stability of the alternating chain in the 
presence of phononic driving, particularly when this is very strong or very 
slow. While self-blocking limits the effectiveness of strong driving, in fact 
the sharp fall in ${\overline n}_{\rm ph0}$ at very small $\omega_0$ in Fig.~1(b) 
of the main text is related to a ground-state instability of the chain, where 
a stimulated distortion can occur ($q_0$ becomes finite) in the presence of 
sufficiently slow phonons. One may show that the stability criterion takes 
the form $\omega_0^c > F(\lambda) g^2 \lambda^2 J$, and that for $\lambda = 
0.5 = g/J$ this critical value is $\omega_0^c \simeq 0.07 J$. 

\section{S2.  Self-blocking}

Here we provide two sets of comments augmenting the discussion of self-blocking 
presented in the main text. Extra insight into the phenomenon is provided if, 
as in Fig.~2(h) of the main text, one considers each of the response functions 
of the spin chain normalized by the actual driving strength at the same 
frequency, which can be gauged from ${\overline n}_{\rm ph0}$. 
In Fig.~\ref{fgsbsm2} we show the normalized spin response, ${\overline 
n}_{\rm x0}$, of Fig.~1(c) of main text. This clearly regains the form of a 
weak-$g$ response \cite{rymfnu} at all $g$, with very strong peaks restored 
at $2 \omega_{\rm min}$ and $2 \omega_{\rm max}$ and the same relative intensities 
at all frequencies. Thus one may state that self-blocking is the only 
significant effect on the relative intensities, and beyond it one observes 
only a minor occupation at the midband energies as strong $g$ spreads the 
resonant response to a wider range of frequencies. It goes without saying 
that this fictitious spin response would be far beyond the physics of the 
spin chain, because of the hard-core nature of the triplon excitations, and 
a different physical mechanism would certainly limit the system to maximally 
nonequilibrium mode populations below $n_{\rm x0} = 1$, but self-blocking 
preempts this by providing a stricter limit. 

\begin{figure}[t]
\includegraphics[width=\columnwidth]{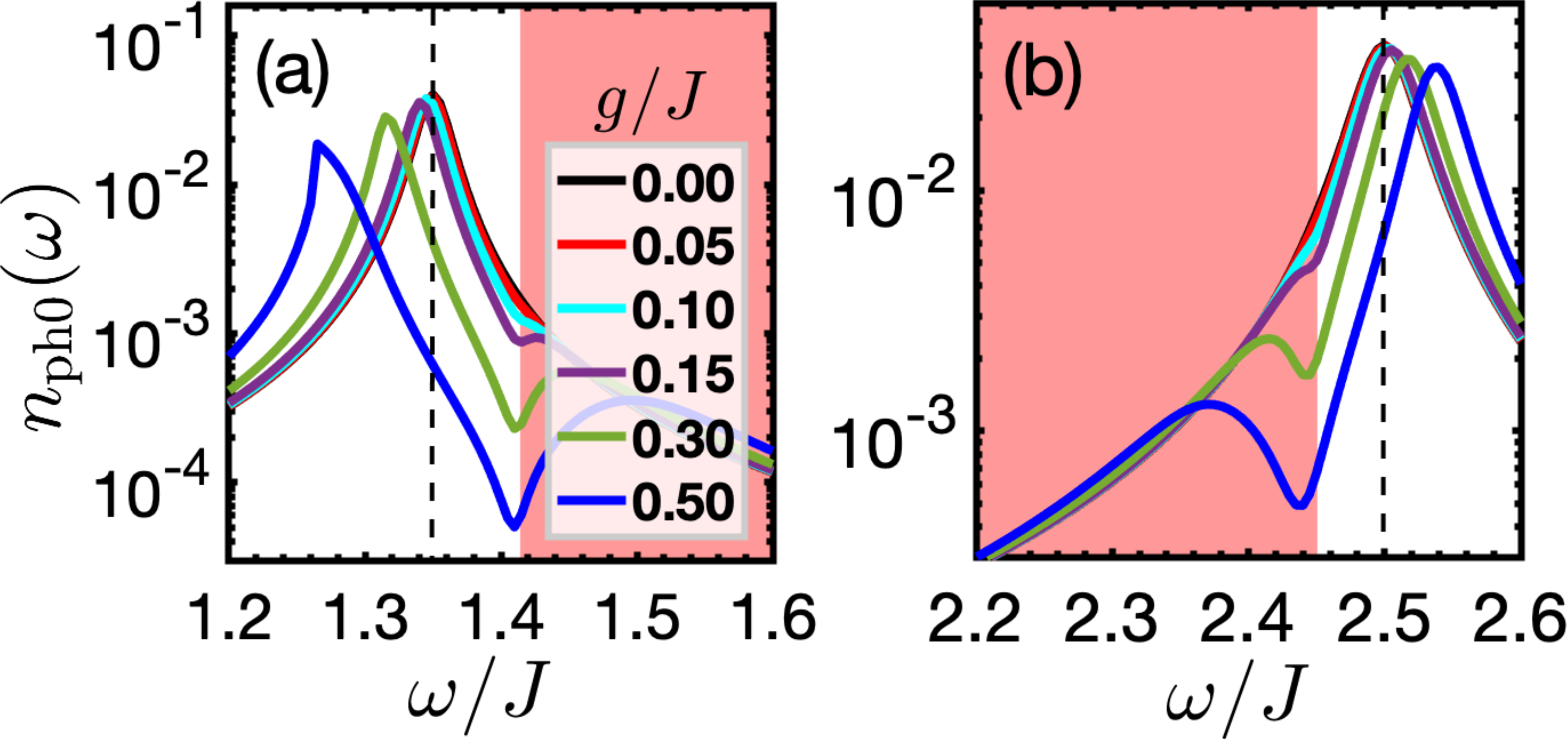}
\caption{(a) Phonon occupation, $n_{\rm ph0} (\omega)$, shown on a logarithmic 
axis for $\omega_0 = 1.35 J$ at selected $g$ values. The standard driving and 
damping parameters of Fig.~1(b) of the main text are used. (b) As in panel (a) 
for $\omega_0 = 2.5 J$.}
\label{fgsbsm3}
\end{figure}

\begin{figure*}[t]
\includegraphics[width=0.96\textwidth]{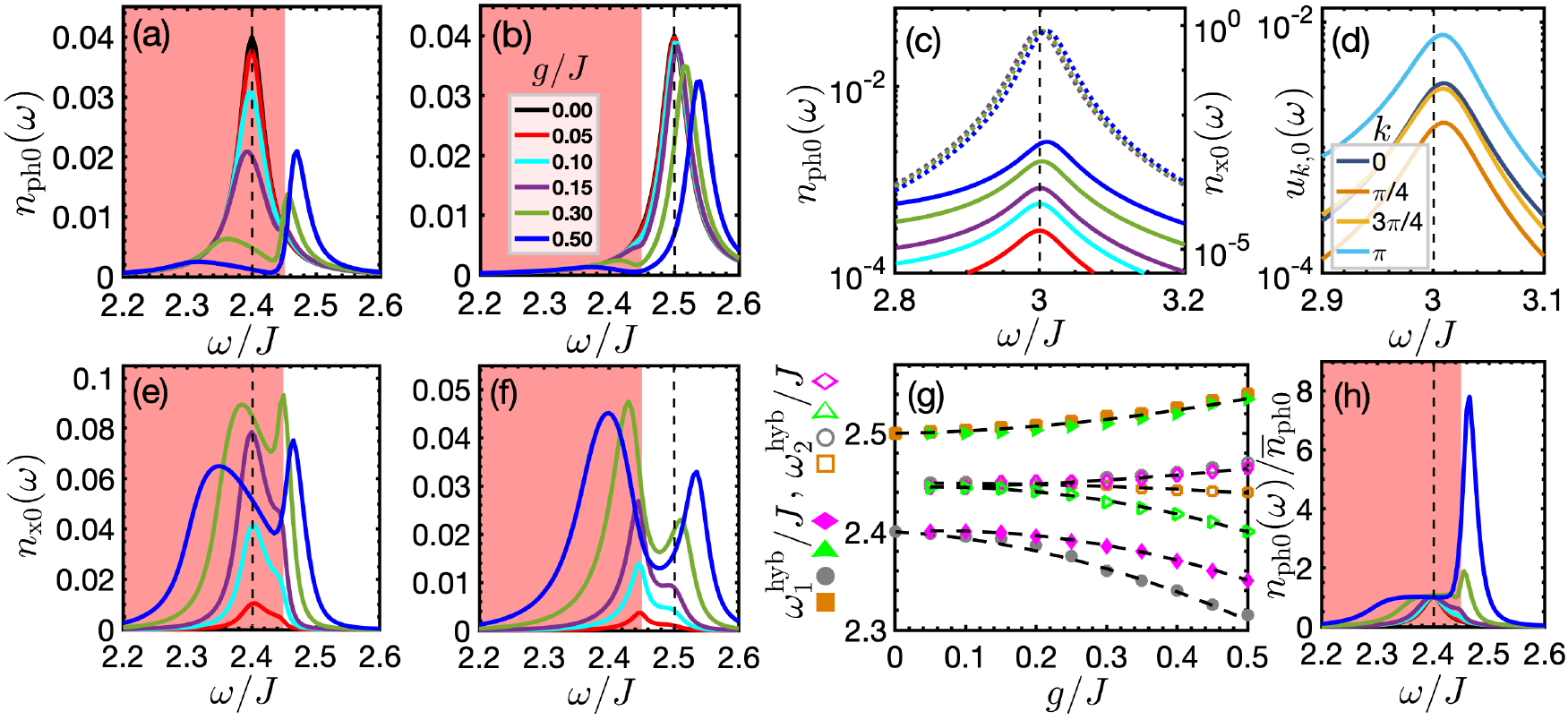}
\caption{(a) Phonon occupation, $n_{\rm ph0}(\omega)$, shown for a phonon 
frequency $\omega_0 = 2.4 J$ at selected $g$ values. The standard driving 
and damping parameters of Fig.~1(b) of the main text are used. 
(b) As in panel (a) for $\omega_0 = 2.5 J$. 
(c) $n_{\rm ph0} (\omega)$ (dotted lines) and triplon occupation, $n_{\rm x0} 
(\omega)$ (solid lines), shown for a phonon frequency, $\omega_0 = 3.0 J$, 
far from the spin excitation band. 
(d) $k$-resolved components of the average $u_k (\omega)$ at $\omega_0 = 3.0 J$.
(e) $n_{\rm x0} (\omega)$ for $\omega_0 = 2.4 J$, corresponding to panel (a). 
(f) $n_{\rm x0} (\omega)$ for $\omega_0 = 2.5 J$, corresponding to panel (b). 
(g) Peak-pair frequencies, labelled $\omega_1^{\rm hyb}$ and $\omega_2^{\rm hyb}$, 
as in panels (a), (b), (e), and (f), shown for all $g$ values; black dashed 
lines indicate a $g^2$ form. 
(h) Data of panel (a) normalized to the phonon occupation ${\overline n}_{\rm 
ph0}$ at the same $\omega$ [data from Fig.~1(b) of the main text].}
\label{fgsbsm4}
\end{figure*}

As noted in the main text, it is also necessary to consider whether the 
self-blocking phenomenon could be a consequence of any simplifying 
assumptions made about the system or the bath. The self-blocking we observe 
is predominantly an effect of feedback on the Einstein phonon, whose coupling 
to the spin system is entirely conventional. Although we illustrated 
self-blocking for a simplified spin bath preserving the independent $k$ 
states of the triplons, even in this situation we demonstrated that 
strong spin-phonon coupling leads to strong shifts of spectral response in 
energy, and hence between modes at different $k$ values. Thus the phenomenon 
is by no means restricted to a single $k$ state at the resonant energy, and 
one may even speculate that a more efficient distribution of spectral weight 
among states of different $k$ could lead to more spin-damping contributions 
(the $\gamma_{\rm s}$ terms of more active modes) and stronger self-blocking. 

Finally, we comment that most ultrafast experiments to date use only 
very short driving pulses \cite{rtkcgms}, which do not fit into the NESS 
framework we have investigated here, but as a consequence do not suffer the 
same heating problems \cite{rymfnu}. They also use very strong electric fields, 
producing instantaneous atomic motion (and hence $n_{\rm ph}$) rather than a 
dependence on the inverse damping times [$\gamma^{-1}$ and $\gamma_{\rm s}^{-1}$ 
in Figs.~\ref{fgsbsm1}(c) and \ref{fgsbsm1}(d)]. Under these circumstances 
one may anticipate that the spin-system feedback ($n_{\rm x}$) also becomes 
instantaneous, and thus that the phenomenon of resonant self-blocking is 
equally applicable to an intensely driven phonon mode coupled strongly to 
a high density of spin states, although its nature in this transient time 
regime remains to be explored. 

\section{S3.  Hybrid excitations}

We augment the analysis of the phonon and spin spectra shown in Fig.~2 of 
the main text by illustrating the ubiquitous nature of the spin-phonon
hybridization phenomena. First, in Fig.~\ref{fgsbsm3}(a) we provide the 
data of Fig.~2(a) of the main text showing $n_{\rm ph0} (\omega)$ on a 
logarithmic scale. In this form it is clear that the ``phononic'' hybrid 
mode being repelled below the band edge with increasing $g$ is accompanied 
by a weak ``magnetic'' hybrid state forming inside the band (although we 
stress again that the figure shows the phononic character of this state). In 
Fig.~\ref{fgsbsm3}(b) we present the analogous data for a phonon just above 
the upper band edge, at $\omega_0 = 2.5 J$, which is the situation we discuss 
next.

\begin{figure*}[t]
\includegraphics[width=0.94\textwidth]{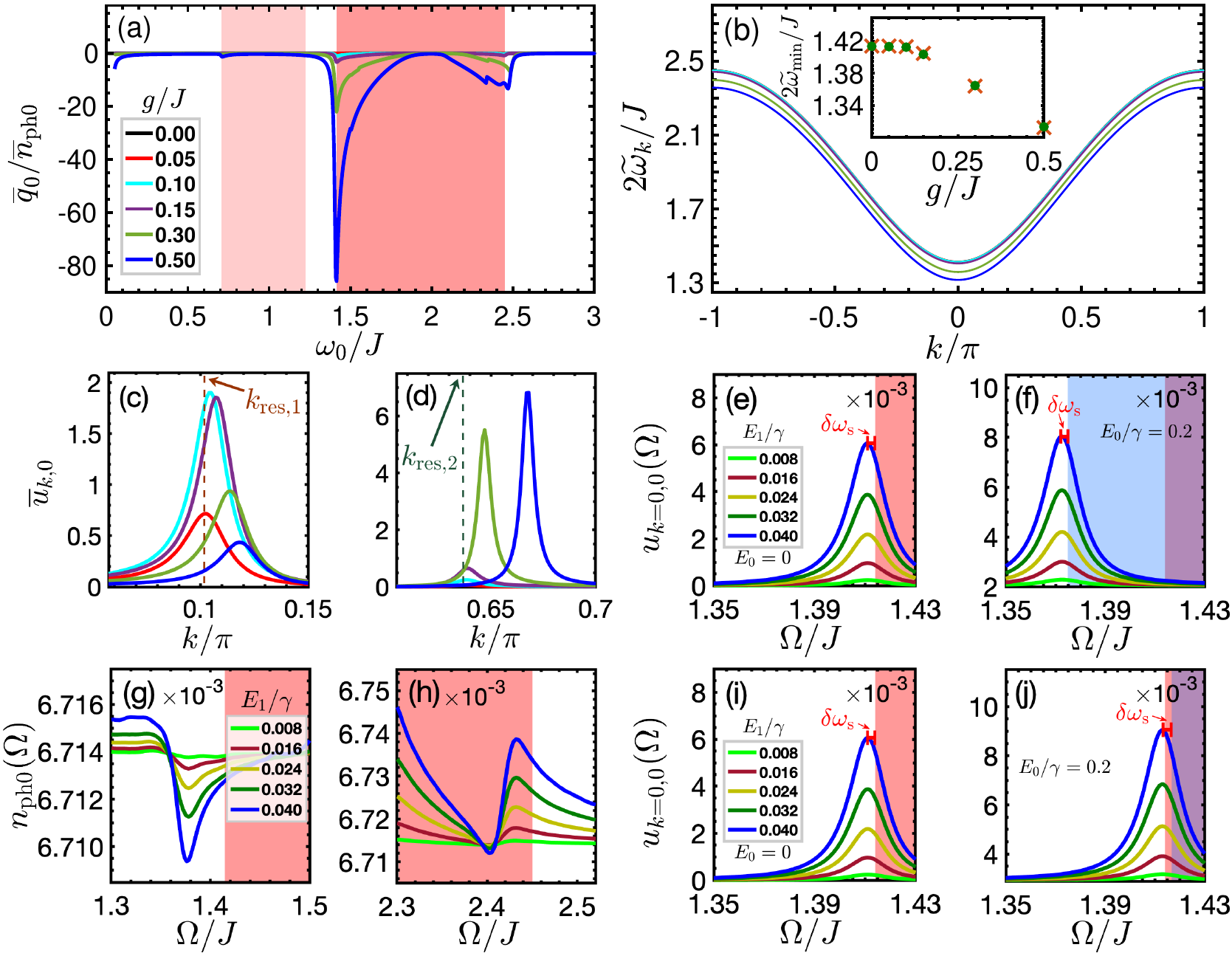}
\caption{(a) Static displacement of the Einstein phonon, ${\overline q}_0$ 
from Fig.~3(a) of the main text, shown with renormalization by the actual 
phonon occupation, ${\overline n}_{\rm ph0}$.
(b) Effective renormalized cosinusoidal triplon dispersion shown in the 
form $2 {\widetilde \omega}_k$, illustrating the extent of the band shifts 
produced by driving with frequency $\omega = \omega_0 = 2.2 J$ and an 
electric field amplitude of twice the standard value. Inset: correspondence 
of the lower band-edge shift (solid circles) with the value obtained from a 
chain with dimer coupling ${\tilde J} = J + g {\overline q}_0$ (crosses). 
(c) Average value of the maximal $u_k$ component, shown for $\omega_0
 = 1.45 J$, where the resonant wave vector (see text) in the isolated 
two-triplon band is $k_{\rm res,1} = 0.102 \pi$. (d) As in panel (c) for 
$\omega_0 = 2.20 J$, where $k_{\rm res,2} = 0.638 \pi$. 
(e) Triplon occupation, $n_{\rm x0} (\omega_0,\Omega)$, obtained at $g = 0.5 J$ 
in the absence of a driving electric field ($E_0 = 0$ at $\omega = \omega_0 = 
2.2 J$) but with a probe electric field, $E_1$, at frequencies $\Omega$ close 
to the lower edge of the two-triplon band, illustrating the persistence of a 
feature at $\delta \omega_{\rm s}$ from the band edge as $E_1 \rightarrow 0$. 
(f) $n_{\rm x0} (\omega_0,\Omega)$ under the same conditions as panel (e) in 
the presence of a driving field, illustrating strong band renormalization and 
the extraction of the renormalized band edge. The blue curve matches the solid 
blue line ($g = 0.5 J$) in Fig.~3(b) of the main text. 
(g) Phonon occupation, $n_{\rm ph0} (\omega_0,\Omega)$, corresponding to the 
driving shown in panel (f) but with stronger probe fields as indicated. 
(h) $n_{\rm ph0} (\omega_0,\Omega)$ corresponding to panel (g), but for 
frequencies $\Omega$ around the upper band edge. 
(i) $n_{\rm x0} (\omega_0,\Omega)$ in the absence of a driving field for a 
model in which the Einstein phonon is coupled to the $J'$ bond of the 
alternating chain [Fig.~1(a) of the main text] with coupling strength 
$g' = g J'/J = 0.5 J'$. 
(j) $n_{\rm x0} (\omega_0,\Omega)$ in the presence of a standard driving field 
for the same model and $g'$.}
\label{fgsbsm5}
\end{figure*}

Figure \ref{fgsbsm4} shows the phononic and magnetic response of the system 
in the situation where the Einstein phonon frequency is close to the upper 
two-triplon band edge. All of the features of Fig.~\ref{fgsbsm4} are at the 
qualitative level symmetrical with those observed in Fig.~2 of the main text, 
which showed the response when $\omega_0$ is close to the lower band edge. 
In Figs.~\ref{fgsbsm4}(a) and \ref{fgsbsm4}(b) we show the development of 
mutually repelling hybrid states as gauged by their phononic response, $n_{\rm 
ph0} (\omega)$, in the presence of a phonon respectively just below or just 
above the band edge. The latter is the situation shown in Fig.~\ref{fgsbsm3}(b),
where in contrast to Fig.~2(a) of the main text the magnetic hybrid is in fact 
discernible (a result reflecting the differences in matrix elements governing 
the two situations). In Figs.~\ref{fgsbsm4}(c) and \ref{fgsbsm4}(d) we show 
the weakly hybridized dressed phononic mode at $\omega_0 = 3.0 J$, where the 
frequency shift due to hybridization is negligible and the weak spin response 
is dominated by $u_{k=\pi,0}$. In Figs.~\ref{fgsbsm4}(e) and \ref{fgsbsm4}(f) 
we show the spin response, $n_{\rm x0} (\omega)$, corresponding to 
Figs.~\ref{fgsbsm4}(a) and \ref{fgsbsm4}(b), noting from the equal heights 
of the peaks on both sides of $2 \omega_{\rm max}$ at all large $g$ values 
that these are again close to the maximally hybridized situation. Figure 
\ref{fgsbsm4}(g) collects all of the characteristic frequencies of the 
phononic and magnetic facets of the strongly hybridized entities, which we 
denote by $\omega_1^{\rm hyb}$ for the in-band peaks in $n_{\rm ph0} (\omega)$ 
and $n_{\rm x0} (\omega)$, and by $\omega_2^{\rm hyb}$ for the above-band peaks. 
As in Fig.~2(g) of the main text, outside (above) the band there is only one 
characteristic frequency for both types of response, whereas inside the 
band there is a not insignificant spread in peak values. Finally, in 
Fig.~\ref{fgsbsm4}(h) we show the phononic response of Fig.~\ref{fgsbsm4}(a) 
normalized by the self-blocking effect of Fig.~1(b) of the main text, which 
emphasizes again the very strong negative feedback effects operating for all 
phonon frequencies within the two-triplon band. 

\section{S4.  Dynamical renormalization}

To gain a more quantitative understanding of how the driving modifies the 
static properties of the coupled system, in Fig.~\ref{fgsbsm5}(a) we show 
${\overline q}_0$ from Fig.~3(a) of the main text normalized by the actual 
phonon occupation, ${\overline n}_{\rm ph0}$, taken from Fig.~1(b) of the main 
text. This allows one to gauge how the band renormalization caused by 
${\overline q}_0$ is in fact controlled completely by self-blocking, becoming 
very small at the band edges, precisely where one might have anticipated the 
strongest effects on the basis of Fig.~\ref{fgsbsm5}(a). 

In Figs.~\ref{fgsbsm5}(c) and \ref{fgsbsm5}(d) we show how the band 
renormalization appears in the pump spectrum, when the system is driven at 
frequency $\omega$ with no probe signal. In this situation one may consider 
the components of $u_{k}$, whose average values peak strongly at the actual 
band frequency, defining a characteristic wave vector, $k_{\rm res}$ 
\cite{rymfnu}. For a phonon at $\omega_0 = 1.45 J$ [Fig.~\ref{fgsbsm5}(c)], 
$u_{{k_{\rm res}},0}$ first undergoes a moderate enhancement due to the effect of 
$g$ before $k_{\rm res}$ is shifted and self-blocking dominates the response. 
For a phonon at $\omega_0 = 2.2 J$ [Fig.~\ref{fgsbsm5}(d)], $u_{{k_{\rm res}},0}$ 
is enhanced and shifted massively at strong $g$, where the downward band shift 
(corresponding to an upward $k_{\rm res}$ shift) is significant and self-blocking 
is weak. However, it is clear from these panels that a quantitative 
characterization of the band renormalization caused by driving phonons 
at frequency $\omega_0$ using a pump at frequency $\omega$ requires the 
introduction of a further frequency, and for this we introduce the probe 
beam at $\Omega$. 

In Figs.~\ref{fgsbsm5}(e) and \ref{fgsbsm5}(f) we illustrate the accurate 
extraction of the new band edge from the probe response. In the absence of 
driving [Fig.~\ref{fgsbsm5}(e)], the response in the probe spectrum peaks 
not at the undriven band edge but at a slightly shifted frequency. This 
shift, $\delta \omega_{\rm s}$ is the result of weak hybridization between 
the triplon pairs near the band edge and the phonon at $\omega_0 = 2.2 J$, 
and its value remains constant as the amplitude of the probe ($E_1$) is 
reduced to zero. When the driving electric field ($E_0$) is restored, in 
Fig.~\ref{fgsbsm5}(f) we observe a large shift in the response peaks and 
we deduce the true location of the renormalized band edge by subtracting 
$\delta \omega_{\rm s}$. 

In Fig.~\ref{fgsbsm5}(b) we show the full cosine band of the triplons to 
illustrate its driven renormalization for different $g$ values. The phononic 
driving takes place at $\omega = \omega_0 = 2.2 J$ and the quantity $2 
{\widetilde \omega}_k$ gives the lower and upper band edges deduced from the 
pump-probe spectra of Figs.~3(b) and 3(c) of the main text when the driving 
electric field is doubled from our standard driving, i.e.~the band edges 
deduced from the dotted lines in these figures. In the inset we compare the 
positions of the lower band edges for each value of $g$, which include the 
relevant $\delta \omega_{\rm s}$ corrections, with the values expected from 
a renormalization of $J$ in Eq.~(\ref{ewk}) to ${\tilde J} = J + g {\overline 
q}_0 (g)$, where ${\overline q}_0(g)$ is taken from Fig.~3(a) of the main text. 
The excellent agreement indicates that any higher-order contributions to the 
driven band renormalization remain small by comparison with the effect of the 
shift in ${\overline q}_0$ for the model we consider [Fig.~1(a) of the main 
text]. 

For further insight into the meaning of our pump-probe results, in 
Fig.~\ref{fgsbsm5}(g) we show the phonon spectrum matching the spin 
response of Fig.~\ref{fgsbsm5}(f), i.e.~for probe frequencies around the 
lower band edge, and in Fig.~\ref{fgsbsm5}(h) the phononic response for 
probing around the upper band edge. The strongest probe fields in these two 
panels match the spin spectra at $g = 0.5 J$ shown in Figs.~3(b) and 3(c) of 
the main text. In the absence of a probe beam, the phonon spectrum is 
essentially flat around the band edges, with no discernible features forming 
in these regions when the only driving is resonant with the available in-band 
phonon at $\omega_0 = 2.2 J$. However, increasing the probe intensity reveals 
that the formation of the predominantly magnetic spectral features at the 
band edge in Figs.~3(b), 3(c), and \ref{fgsbsm5}(f) is accompanied by a small 
dip in $n_{\rm ph0} (\omega_0,\Omega)$. This weak response indicates the weak 
phononic character of these hybrid states; the fact that it is negative is 
again a consequence of self-blocking, in that the resonant hybrid interferes 
weakly with the uptake of laser energy by the phonon.

\section{S5.  $J'$ Model}

To gain further perspective on the nature of magnetophononic self-blocking 
and triplon band engineering, we consider a model in which the Einstein 
phonon is coupled to the (interdimer) $J'$ bond in Fig.~1(a) of the main 
text, i.e.~the spin-phonon part of the Hamiltonian in Eq.~(\ref{ehp}) is 
changed to
\begin{equation}
H_\text{sp} = \sum_j g (b_j + b_j^\dag) \, {\vec S}_{2,j} \! \cdot \! {\vec 
S}_{1,j+1}.
\label{ehpjp}
\end{equation}
This ``$J'$ model'' has two primary differences from the ``$J$ model'' 
considered hitherto. First, because the spin-phonon coupling has dimensions 
of $J$ but the interdimer bond has magnitude $J'$, the effect of the coupling 
term [Eq.~(\ref{ehpjp})] is amplified in the equations of motion and it is 
convenient to compare the $J$-model results to a $J'$ model with the rescaled 
coupling $g' = g J'/J$. Second, the coefficients in Eqs.~(\ref{eyk}) 
and (\ref{eypk}) undergo the alterations $y_k \to - y_k'/ \lambda$ and $y_k' 
\to - y_k/\lambda$, which lead to specific changes in the physics. 

First for self-blocking, if $g$ is replaced by $g'$ ($= g/2$ for the 
illustrative parameters we use) then the results for ${\overline n}_{\rm ph0}$ 
are numerically very close to those of Fig.~1(b) of the main text. This 
reinforces our statement that self-blocking is indeed a generic phenomenon 
in a driven spin-phonon-coupled system, rather than possibly being a special 
consequence of localization, dimerization, and unit-cell selection. Turning 
to hybrid-state formation, the phenomenology of Figs.~2 and \ref{fgsbsm4} 
is also reproduced in the $J'$ model, with frequency shifts that are 
approximately four times as large if $g$ is not replaced by $g'$. 

Finally, triplon band engineering in the $J'$ model is quite different because 
of the altered coefficients mentioned above: ${\overline q}_0$ is smaller by 
one order of magnitude than the values found in Fig.~3(a) of the main text, 
meaning that the band-shifts of Figs.~3(b) and 3(c) are not attainable. 
However, ${\overline q}_0$ also changes sign as the driving phonon frequency 
passes through the center of the two-triplon band ($\omega_0 = 2J$), indicating 
that phonons coupled to $J'$ offer a different type of band control, in the 
form of a band-narrowing or -broadening. These results are not surprising 
if one considers the expression for $\omega_k$ in Eq.~(\ref{ewk}). Figure 
\ref{fgsbsm5}(i) shows the probe response at the lower band edge of 
an undriven model with $g' = 0.5 J'$, from which we observe that the 
frequency shift, $\delta \omega_{\rm s}$, is almost identical to that in 
Fig.~\ref{fgsbsm5}(f). In Fig.~\ref{fgsbsm5}(j) we observe that the (upward) 
shift of the lower band edge caused by driving this model at $\omega_0 = 2.2 
J$ is very weak, and in fact it corresponds to a band-narrowing (the upper 
edge is pulled downwards). We comment for completeness that driving a phonon 
at $\omega = \omega_0 = 1.8 J$ gives the strongest response in the $J'$ model, 
a band-broadening that is approximately twice as large for the same driving 
strength. 

\end{document}